\documentstyle[12pt]{article}
\begin{document}

\vspace{1cm}

\begin{center}
{\bf WHAT DO WE LEARN FROM ATOMIC PHYSICS }
\end{center}

\begin{center}
{ABOUT FUNDAMENTAL SYMMETRIES?}
\end{center}

\vspace{1cm}

\begin{center} 
Invited talk at Workshop on Parity and Time Reversal Violation
\end{center}
\begin{center}
in Compound Nuclear States and Related Topics,
\end{center}
\begin{center}
Trento, Italy, October 16-27, 1995
\end{center}

\vspace{1cm}

\begin{center}
{\bf I.B. Khriplovich }\footnote{e-mail address: khriplovich@inp.nsk.su}
\end{center}
\begin{center}
Budker Institute of Nuclear Physics, 630090 Novosibirsk, Russia
\end{center}

\vspace{2.0cm}

\begin{abstract}
Valuable information on interactions violating $P$- and $T$-invariance
can be extracted from atomic experiments. 
The hypothesis of a large weak matrix element between single-particle
states in heavy nuclei, $\sim 100$ eV, is ruled out by measurements of
parity nonconservation in atoms.
Experimental upper limit on the electric dipole moment of 
the $^{199}$Hg atom strictly constrains  parameters of $CP$-violation
models.
Upper limit on the $T$-odd, $P$-even admixture to nuclear forces 
is set: $\alpha_T\,<\,10^{-11}$.
\end{abstract}

\vspace{2cm}

\section{Is Large Weak Mixing in Heavy Nuclei Consistent 
with Atomic Experiment?}

The scattering cross-sections of longitudinally polarized epithermal
(1 -- 1000 eV) neutrons from heavy nuclei at $p_{1/2}$ resonances have
large longitudinal asymmetry. This parity nonconserving (PNC)
correlation is the fractional difference of the resonance
cross-sections for positive and negative neutron helicities. For a
long time the most natural explanation of the effect was based on the
statistical model of the compound nuclei. In fact, not only the
explanation, but the very prediction of the huge magnitude of this
asymmetry (together with pointing out the nuclei most suitable for the
experiments) was made theoretically \cite{sf} on the basis of this
model.

An obvious prediction of the statistical model is that after
averaging over resonances, the asymmetry should vanish. However, few
years ago it was discovered \cite{fb,fb1} that all seven asymmetries
for $^{232}$Th have the same, positive sign. 

Most attempts \cite{bg,kj,ab,lw} to explain a common sign, require
the magnitude of the weak interaction matrix element, mixing
opposite-parity nuclear levels, to be extremely large, $\sim 100$ eV. 
The same assumption seems to be necessary to explain unexpectedly
large $P$-odd correlations observed in the M\"{o}ssbauer transitions in
$^{119}$Sn and $^{57}$Fe \cite{ts,ts1}.

Such a large
magnitude of the weak mixing can be checked in independent
experiments. The recent proposal \cite{sb} 
is to measure PNC asymmetry in the M4
$\gamma$-transition between the 
states $1i\; 13/2^+$ and $2f\; 5/2^-$ in $^{207}$Pb, 
which are believed to be predominantly single-particle
ones. The experiment
is in progress now \cite{sb1}. Its sensitivity to the weak matrix 
element value is expected to reach 
$5 - 13$ eV. 

In \cite{dkt1} it was demonstrated that close upper
limit on the weak mixing in $^{207}$Pb follow already from the
measurements of the PNC optical activity of atomic lead vapour
\cite{mvm}. The experiment was performed at the atomic M1 transition
from the ground state $6p^2\;^3P_0$ to the excited one $6p^2\;^3P_1$.
The nuclear spin of $^{207}$Pb being $i=1/2$, the total atomic angular
momentum of the ground level is $F=1/2$, and the upper level is split
into two: $F^{\prime}=1/2,\,3/2$. The following upper limit was
established at the 95\% confidence level for the relative magnitude of
the nuclear-spin-dependent (NSD) part of the optical activity:
\begin{equation}\label{ra}
\frac{P_{NSD}}{P} < 0.02  
\end{equation}
Here 
$$P_{NSD} 
= P(F=1/2\rightarrow F^{\prime}=1/2) - P(F=1/2\rightarrow F^{\prime}=3/2)$$
and $P$ is the main, nuclear-spin-independent, part of the PNC
optical activity.

In heavy atoms the NSD $P$-odd effects were shown to be induced mainly
by contact electromagnetic interaction of electrons with the anapole
moment of a nucleus, which is its $P$-odd electromagnetic characteristic
induced by PNC nuclear forces \cite{fk,fks1,khr}.

The electromagnetic PNC interaction of electrons with nuclear AM is
of a contact type. It is conveniently characterized in the units of
the Fermi weak interaction constant $G=1.027\times 10^{-5} m^{-2}$
($m$ is the proton mass) by a dimensionless constant $\kappa$. 

To calculate, $\kappa$ we present the effective $P$-odd potential
for an external nucleon in a contact form in the spirit of the
Landau-Migdal approach:
\begin{equation}\label{we}
W=\frac{G}{\sqrt{2}}\;\frac{g}{2m}\;{\bf \sigma}[{\bf p}\rho(r)
+\rho(r){\bf p}\;].
\end{equation}
Here ${\bf \sigma}$ and ${\bf p}$ are respectively spin and momentum
operators of the valence nucleon, $\rho(r)$ is the density of
nucleons in the core normalized by the condition $\int
d{\bf r}\rho(r)=A$ (the atomic number is assumed to be large, $A\gg
1$). A dimensionless constant $g$ characterizes the strength of the
$P$-odd nuclear interaction.  It is an effective one and includes
already the exchange terms for identical nucleons. This constant
includes also additional suppression factors reflecting long-range
and exchange nature of the $P$-odd one-meson exchange, as well as the
short-range nucleon-nucleon repulsion.

Under some simplifying assumptions the anapole constant $\kappa$ can
be estimated for a heavy nucleus even analytically with the following
result \cite{fks1}:
\begin{equation}
\kappa=\frac{9}{10}\,g\,\frac{\alpha \mu}{m r_0}\,A^{2/3}.
\end{equation}
Here $\mu$ is the outer nucleon magnetic moment, $r_0=1.2\,$fm. The
enhancement $\sim A^{2/3}$ compensates to a large extent the small
fine structure constant $\alpha=1/137$. That is why the nuclear AM is
perhaps the main source of the nuclear-spin-dependent PNC effects in
heavy atoms \cite{fk,fks1}. This formula predicts for lead
\begin{equation}
\kappa(^{207}Pb) = -\,0.08\;g_n.
\end{equation}
More serious numerical calculations using a realistic description of
the core density and the Woods-Saxon potential including the spin-orbit
interaction, give \cite{fks1,dkt} 
\begin{equation}
\kappa(^{207}Pb) = -\,0.105\; g_n.
\end{equation}
Recently it was demonstrated\cite{dt} that various many-body corrections,
taken together, do not change essentially this result.

On the other hand, atomic calculations predict the magnitude of the
NSD optical activity in lead at given $\kappa$ with the accuracy
about 20\% \cite{nsfk,k}. At the experimental value of $P$ obtained
in \cite{mvm} this prediction for the ratio (\ref{ra})
constitutes 
\begin{equation}
0.023\, \kappa(^{207}Pb).
\end{equation}
Combining the experimental result (\ref{ra}) with this theoretical one,
we get the following upper limit for the anapole constant:
\begin{equation}
\kappa(^{207}Pb) \,<\, 1,
\end{equation}
and for the effective neutron PNC constant:
\begin{equation}\label{co}
g_n \,<\, 10.
\end{equation}

Close upper limits on the effective constant $g_p$ for an outer
proton can be extracted from the optical experiments with atomic
cesium \cite{wi} and thallium \cite{baird,la}. Less strict bound on $g_p$
follows from the experiment \cite{mzw} with bismuth. 

A simple-minded estimate for the weak mixing matrix element, based on
formula (\ref{we}), leads to its following value: 
$$2\,g\;\mbox{eV}.$$
More sophisticated calculations based on the Woods-Saxon potential with
the spin-orbit interaction, give for the concrete matrix element of
interest for the proposed experiment with $^{207}$Pb
\begin{equation}\label{ca}
\langle 3d\; 5/2^+|W| 2f\; 5/2^-\rangle = 1.4\,g_n\;\mbox{eV}
\end{equation}
in a reasonable agreement with the results of other single-particle
nuclear calculations cited in \cite{sb}. Combining (\ref{co})
and (\ref{ca}), we get the following upper limit on this matrix
element
\begin{equation}
\langle 3d\; 5/2^+|W| 2f\; 5/2^-\rangle < 14 \;\mbox{eV}
\end{equation}
which is close to the expected accuracy of the experiment discussed
in \cite{sb,sb1}. 

Nevertheless, this experiment would be obviously
both interesting and informative.
As to the hypothesis itself, according to which the value of the weak
mixing matrix element is as high as 100 eV, it does
not agree with the results of the atomic PNC experiments.

\section{$CP$-Violation without Strangeness, \newline
or Electric Dipole Moments}

Up to now $CP$-violating effects have been observed only in the decays
of the neutral $K$-mesons, and their nature remains mysterious. At present
the searches for the electric dipole moments (EDM) in neutron and atomic 
physics are practically the only other source of the information on
$CP$-violation. Even without discovering the effects sought for, the 
neutron and atomic experiments have ruled out most models of $CP$-violation  
suggested to explain the effects in $K$-meson decays; in fact, one can
argue that the neutron EDM experiments have ruled out more theoretical
models than any other experiment in the history of physics. As to the 
mechanism of $CP$-violation incorporated into the standard model of 
electroweak interactions, which is most popular at present, its predictions
for the EDMs are many orders of magnitude below the present experimental
bounds.

But does it mean that the EDM experiments are of no serious interest for the
elementary particle physics, are nothing else but mere exercises in
precision spectroscopy? Just the opposite. It means that these 
experiments now, 
at the present level of accuracy are extremely sensitive to possible 
new physics beyond the standard model, physics to which the kaon decays are
insensitive. Examples of this type will be discussed in more detail below.

\subsection{Phenomenological $CP$-Odd Nuclear Potential \newline
and the Schiff Moment} 

The most strict upper limit on the EDM of anything is obtained in the
recent experiment \cite{lam} with $^{199}$Hg atom:
\begin{equation}\label{hg}
d(^{199}\mbox{Hg})/e\,<\,9\cdot 10^{-28}\,\mbox{cm}.
\end{equation}
The electronic shells of this diamagnetic atom are closed. Therefore, the
effect is due to the nuclear EDM. To be more precise, because of the
electrostatic screening, the atomic 
dipole moment (\ref{hg}) is induced 
not by the nuclear EDM itself, but by the so-called Schiff moment, which is 
proportional to the difference between the nuclear dipole moment form-factor
and nuclear charge form-factor (see, e.g., book \cite{khr}).

It has been demonstrated in \cite{sfk} that nuclear $CP$-odd
electromagnetic moments, the Schiff moment included, are most 
efficiently induced by $T$- and $P$-odd
nuclear forces, but not by the nucleon EDM.

So, let us start with discussing the $T$- and $P$-odd nucleon-nucleon
potential. If we assume for simplicity the interaction to be local
and limit ourselves to first-order terms in the nucleon velocities
$p/m_p$, then to this approximation the most general form of the
effective potential (in the spirit of the Landau-Migdal approach) is
\[W_{ab}=\,\frac{G}{\sqrt{2}}\;\frac{1}{2m}\;\left\{(\xi_{ab}
{\bf {\sigma}}_a
\,-\,\xi_{ba}{\bf {\sigma}}_b)\,
{\bf {\nabla}}\,\delta({\bf r}_a-\,{\bf r}_b)\right.\]
\begin{equation} \label{wab}
\left.+\,\xi_{ab}^{\prime}[{\bf \sigma}_a\times{\bf \sigma}_b]
\{({\bf p}_a-{\bf p}_b)\,\delta({\bf r}_a-\,{\bf r}_b)\,+\,
\delta({\bf r}_a-\,{\bf r}_b)\,({\bf p}_a-{\bf p}_b)\}\right\}
\end{equation}
The dimensionless constants $\xi$, characterizing the strength of the
interaction in units of the Fermi constant $G$, are supplied with
subscripts in order to distinguish between protons and neutrons. These
are effective constants and include already the exchange terms for
identical nucleons. In a detailed theory the constants should also
include additional suppression factors reflecting long-range and
exchange nature of the realistic interaction, as well as the
short-range nucleon-nucleon repulsion. As it should be, the
potential (\ref{wab}) is invariant under the Galilean
transformations. 

According to detailed numerical calculations  
with the Woods-Saxon potential,
including the spin-orbit interaction, the Schiff moment of $^{199}$Hg
nucleus is \cite{fks}:
\begin{equation}\label{S}
S(^{199}\mbox{Hg})/e\,=\,-\,1.8\cdot 10^{-7}\,\xi_{np}\;\mbox{fm}^3.
\end{equation}
On the other hand, the Hartree-Fock calculations relate the mercury 
atomic dipole moment $d(^{199}\mbox{Hg})$ to the nuclear Schiff moment 
$S(^{199}\mbox{Hg})$ as follows \cite{amar,fks}:
\begin{equation}\label{d}
d(^{199}\mbox{Hg})\,=\,-\,3\cdot 10^{-18}
\left(\frac{S(^{199}\mbox{Hg})}{\mbox{fm}^3}\right)\,\mbox{cm}.
\end{equation}
Combining relations (\ref{S}) and (\ref{d}) with the experimental result
(\ref{hg}), we obtain the following upper limit on the $CP$-odd nuclear
interaction:
\begin{equation}\label{xi}
\xi_{np}\, <\, 1.7\cdot 10^{-3}.
\end{equation}

There are strong reasons to believe that the
$CP$-odd $NN$ interaction in nuclei is dominated  by the
$\pi^0$-exchange.  This mechanism, 
considered first in \cite{hh1}, is singled
out by the large value,
13.6, of the strong $\pi NN$ constant and by the small $\pi$-meson
mass. The derivative occurring at one of the vertices (in this case at the
strong one) arises inevitably in the case of a $P$-odd interaction
and does not lead to a relative suppression of the corresponding
contribution. Finally, a charged particle exchange is suppressed as
compared to neutral exchange in the nuclear shell model.

To simplify the discussion, we will confine ourselves to the limit of
zero momentum transfer in the $NN$ interaction. In this way the
$\pi^0$-exchange induces an effective operator
\begin{equation}\label{NN}
\frac{G}{\sqrt 2}\,\xi\,(\bar N\,i\gamma_5 N)(\bar{N}^{\prime}N^{\prime})
\end{equation}
with a dimensionless constant 
\begin{equation}
\xi\,=\,\frac{g_{\pi NN}\,\bar{g}_{\pi NN}\sqrt 2}{G\,m_{\pi}^2}.
\end{equation}
In the nonrelativistic reduction, expression (\ref{NN}) generates in
the coordinate representation interaction (\ref{wab}).
Then the result (\ref{xi}) 
can be formulated as an upper limit on the effective
$CP$-odd neutral pion constant:
\begin{equation}\label{gex0}
\bar{g}_{\pi NN}^0\,<\,2 \cdot 10^{-11}.
\end{equation} 

\subsection{Quark Chromoelectric Dipole Moment \newline 
and Constraints on Models of $CP$-violation}

The last upper limit is most efficiently employed for constraining models of
$CP$-violation in the following way. Let us consider the effective operator
of the $CP$-odd quark-gluon interaction
\begin{equation}\label{c}
H_{c}=\,\frac{1}{2}\,d^c\,\bar q \gamma_5\sigma_{\mu\nu}t^a q\,G^a_{\mu\nu}.
\end{equation}
where $t^a\,=\,\lambda^a/2$ are the generators of the colour $SU(3)$
group. This is a close analogue of the EDM interaction with the 
electromagnetic field, so it is only natural to call the constant $d^c$ 
in expression (\ref{c}) the
quark chromoelectric dipole moment (CEDM). The $CP$-odd $\pi^0 NN$ vertex
generated by operator (\ref{c}) transforms by the PCAC technique: 
\begin{equation}\label{pi0}
<\,\pi^0 N\,|\,g\,\bar q \gamma_5\sigma_{\mu\nu}t^a
q\,G^a_{\mu\nu}|\,N\,> =\,\pm \, \frac{i\sqrt 2}{f_{\pi}}\,
 <\,N\,|\,g\,\bar q \sigma_{\mu\nu}t^a q\,G^a_{\mu\nu}\,|\,N\,>; 
\end{equation}
the plus and minus in the lhs refer to the $u$- and $d$-quark CEDM,
respectively. The QCD sum rule estimate for the last expectation value is
$7\,\mbox{GeV}^2\;$ \cite{kky}. Combining it with (\ref{gex0}), 
we obtain the upper limit for the quark CEDM:
\begin{equation}
d^c\,<\,2.4\cdot 10^{-26}\,\mbox{cm}.
\end{equation}

Let us consider now the model of spontaneous $CP$-violation
in the Higgs sector. Its old version, with light Higgs bosons, has
been ruled out by the experimental upper limit on the neutron EDM.
We will consider therefore its more ``natural" version, with heavy Higgs
bosons. Of course, in this case the model is
responsible for only a small portion of $CP$-violation in kaon decays. 
It would be new physics, a new source of $CP$-violation, supplemental
to that generating the effects already observed.

The estimate for the quark CEDM obtained in the ``natural" version 
of the model, under
the assumption that the Higgs mass is about the same as that of the
$t$-quark, is \cite{gw,cky}:
\begin{equation}\label{dcq}
d^c(q) \sim 3\cdot 10^{-25}\,\mbox{cm}.
\end{equation}
As it is pointed out in \cite{kh1u}, this prediction is
12 times larger than the experimental upper limit
(\ref{gex0}). 

Thus, very special assumptions concerning the parameters of the model
of spontaneous $CP$-violation in the Higgs sector (such as large mass
of the Higgs boson, small values of the $CP$-violating parameters,
etc) are necessary to reconcile the predictions of this model with
the experimental upper limits on the neutron electric dipole moment.

The same situation takes place in the
supersymmetric $SO(10)$ model \cite{khz}. 

Let us mention here that 
the atomic experiment constrains the parameters of both model
more strongly than the upper limit on the neutron EDM.

\section{What Do We Know in Fact 
about $T$-odd, \newline 
but $P$-even interactions?}

Direct experimental information on the $T$-odd, $P$-even (TOPE)
interactions is rather poor. Best limits on the relative magnitude of
the corresponding admixtures to nuclear forces lie around $10^{-3}\;\;$
\cite{chk,blh,dav,fps}. We will relate below all interactions to the
Fermi weak interaction constant $G$.  Since the nuclear scale of weak
interactions is $Gm^2_{\pi}\sim 2 \cdot 10^{-7}$, those limits can
be formulated as $10^4\, G$. Most advanced experimental experimental
proposals aim at improving these limits by three orders of magnitude.

Experimental information on TOPE electron-nucleon interaction is
practically absent. In \cite{porkoz} an atomic experiment was
suggested which can hopefully reach an accuracy about $\sim 3 \cdot
10^4 G$ (see also \cite{mospor}). Higher accuracy is aimed at in
the recent experimental proposal \cite{cont}.

As to the TOPE electron-electron interaction, its possible manifestations
in positronium were discussed in \cite{chr}.

Much better upper limits on the TOPE interactions can be obtained as
follows. Radiative corrections, due to the $P$-odd part of the
electroweak interaction, transform the $T$-odd, but $P$-even
fermion-fermion interaction into a $T$-odd and $P$-odd one. The
experimental information about $T$-odd, $P$-odd effects is
sufficiently rich to obtain in this way new limits, much better than
direct ones, on the parameters of $T$-odd, $P$-even
electron-electron, electron-nucleon and nucleon-nucleon interactions.

\subsection{Long-Distance Effects}

Let us point out first that the predictions of all modern
renormalizable theories of $CP$-violation (and not only the standard
model!) cannot exceed $(10^{-3} - 10^{-4})\,G$. The reason is obvious.
Parity violation is an intrinsic property of all these models, and
therefore $T$-odd, $P$-even effects should be roughly of the same order
of magnitude as $T$-odd, $P$-odd ones. 

An even stronger result was obtained recently in \cite{hksw}. In any
renormalizable theory, TOPE flavour conserving quark-quark
interactions are absent to second order in the electroweak coupling.
This conclusion holds to all orders both in the chromodynamic and
electromagnetic interactions, if the $\theta$-term is neglected.

Therefore, the investigations in this field are in fact the search 
for an essentially new
physics, well beyond the modern theories. That is why we will describe the 
TOPE interactions phenomenologically, using effective quark-quark
operators.

There is only one (up to the interchange $1\leftrightarrow 2$) such
operator \cite{tkh1}, which can be presented as
\begin{equation}\label{da}
\frac{G}{\sqrt 2}\,\frac{q_1}{2m}\,
\bar{\psi}_1i\gamma_5 \sigma_{\mu\nu}(p'_1-p_1)_\nu \psi_1
\bar{\psi}_2\gamma_\mu\gamma_5\psi_2.
\end{equation}
We measure the interaction discussed in the
units of the Fermi weak interaction connstant $G$; 
the choice of $m$ as the necessary dimensional parameter being also a
matter of convention; $q_{1}$ is dimensionless.

A hint at the kind of limits that can be obtained by combining this
interaction with the $P$-odd one, is given by the following
argument, close in spirit to the corresponding estimates 
from \cite{wol,her}. (From now on we sacrifice the purity of style, 
and use freely the results of neutron experiments in line with
atomic ones.) Let us consider the contribution to the neutron EDM
from the combined action of the usual $P$-odd,
$T$-even weak interaction and the discussed $T$-odd and $P$-even
interaction, the strength of the latter being $q$ times smaller than
that of the previous one. The contribution constitutes obviously
\begin{equation}\label{wh}
d(n)/e\,\sim\,\frac{1}{m_p}(Gm^2_\pi)^2 q\; 
\end{equation}
From the comparison with the experimental upper limit for the neutron
EDM \cite{sm,al}
\begin{equation}\label{nedm}
d(n)/e \,<\,10^{-25}\,\mbox{cm},
\end{equation}
we obtain the
limit $q<10^2$, which is about two orders of magnitude better than
the direct limits mentioned above. This estimate is obviously of a
very crude nature. In particular, the dipole moment arises here at
least in one-loop approximation which leads to a small geometrical
factor. So, it is better perhaps to accept for this limit a more
cautious estimate
\begin{equation}\label{cru}
q\,<\,10^2-\,10^3,
\end{equation}
which can be otherwise formulated as an upper limit for the 
relative magnitude $\alpha_T$ of the TOPE admixture to nuclear forces:
\begin{equation}
\alpha_T\,<\,10^{-5}-\,10^{-4}.
\end{equation}

Analogous estimates for the electron-nucleon interaction were made in
\cite{kozl,khrip} (as cited in \cite{tkh1}) and \cite{step}. Let us
mention also recent elaborate investigations \cite{hh,hhm,egh,vor} $\;$ 
of
the long-distance interplay between TOPE and usual $P$-odd
interactions in the hadronic sector. They are of a certain interest
for the theoretical nuclear physics, but none of them resulted in a
serious improvement over the simple-minded estimate (\ref{cru}).

Better limits on TOPE
effects are obtained by a simple and 
elegant argument presented in a recent paper \cite{efs}. By 
dimensional reasons, the
TOPE 4-fermion effective interaction of dimension seven can be
written as 
\begin{equation}
\frac{C_7}{\Lambda^3}\,\bar{\psi}_1i\gamma_5 \sigma_{\mu\nu}
(p'_1-p_1)_\nu \psi_1
\bar{\psi}_2\gamma_\mu\gamma_5\psi_2.
\end{equation}
It is only natural to assume that the momentum scale $\Lambda$ in
this operator exceeds that of the electroweak theory, i.e.,
\[\Lambda\,>\,100 \,\mbox{GeV}.\]
As to the dimensionless number $C_7$, it is natural to assume
that it is about unity. 
Then the dimensional estimate for the magnitude of TOPE effects on
the hadronic scale of momenta $p\,\sim\,1$ GeV, is
\begin{equation}
\left(\frac{p}{\Lambda}\right)^3\,<\,10^{-6}.
\end{equation}

This line of reasoning applies not only to 4-fermion 
operators, but, for instance, to a quark-gluon-photon operator of the
form \cite{efs}
\begin{equation}  
\frac{C_7'}{\Lambda^3} \bar{q}
\sigma_{\mu\nu} t^a q\,G_{\mu\rho}^a F_{\nu \rho}. 
\end{equation}
Let us mention here also the TOPE
photon-fermion scattering amplitudes. They belong to higher
dimensions (starting from 10) \cite{tkh4}, and are  
constrained more strongly in this way.
 
\subsection{TOPE Fermion-Fermion Interactions. \newline
One-Loop Approach}

A serious advance in the problem is due to the observation 
that the electroweak
corrections to TOPE fermion-fermion operators are controlled 
mainly not by the
large-distance effects, but by short-distance ones. Therefore,
they are of the order $\alpha/\pi$ (up to some chiral suppression
factor which is quite essential), but not of the order $Gm^{2}_\pi$
\cite{tkh1}. 

We will concentrate here and below on the corrections due to the
$Z$-boson exchange. These can be calculated self-consistently
in the sense that the result is independent of the choice of the
gauge for the $Z$-boson propagator. 

A consistent, gauge-independent calculation of the $W$-boson exchange
contribution to the induced $T$- and $P$-odd amplitudes is much more
model-dependent and will not be discussed here in detail. It can be
expected however to be even larger than that of the $Z$-exchange, due
to small numerical values of the neutral weak charges. These small
values are responsible in particular for the well-known relative
suppression of the neutral-current cross-sections as compared to the
charged-current ones.  So, the $Z$-exchange contribution serves as an
estimate from below for the effects discussed. As to the Higgs boson
exchange, in the standard model it conserves parity and is therefore
of no interest to us.

The result of the transformation of this effective operator into
$T$- and $P$-odd one is
\begin{eqnarray}
\frac{G}{\sqrt 2}\,\frac{\alpha}{3\pi}\, \log \frac{\Lambda^2}{M^2}\, 
\frac{q_1}{m}\,
\{\, 2 m_2 v_1\,[\,3 a_2 \bar \psi_1 \psi_1\,\bar \psi_2\, i \gamma_5 \ \psi_2 
-\,(a_1 + a_2)\bar \psi_1\psi_1 \bar \psi_2 i \gamma_5 \psi_2\,] 
\nonumber \\
  - a_1v_2\,[\,m_2 \bar \psi_1 i \gamma_5 \sigma_{\mu \nu} \psi_1\,
\bar \psi_2 \sigma_{\mu \nu} \psi_2\, - \bar \psi_1 i \gamma_5
\sigma_{\mu \nu}(p'_1 - p_1)_\nu \psi_1\,
\bar \psi_2 \gamma_\mu \psi_2\,]\, \}.
\end{eqnarray}
Here $M$ is the $Z$-boson mass. The dependence of the result 
on the cut-off parameter $\Lambda$ is due to nonrenormalizability 
of the TOPE interaction. But trying to be as conservative as possible
in our numerical estimates, we will assume the log to be of the order
of unity. $m_{1,2}$ are the masses of the first and second fermions,
respectively.  $v_{1,2}$ and $a_{1,2}$ are their weak neutral vector
and axial charges. In particular, for the electron, $u$-, and
$d$-quarks they are:
\begin{eqnarray}\label{ch}
v_{e}=-\frac{1}{2}(1-4\sin^2\theta)\approx -0.04,                 &  a_{e}=
-\frac{1}{2},\nonumber\\
v_{u}=\frac{1}{2}(1-\frac{8}{3}\sin^2\theta)\approx\frac{1}{6},   &  a_{u}=
\frac{1}{2},\nonumber\\
v_{d}=-\frac{1}{2}(1-\frac{4}{3}\sin^2\theta)\approx-\frac{1}{3}, &  a_{d}=
-\frac{1}{2}.
\end {eqnarray}
Here $\theta$ is the electroweak mixing angle,
\begin{center}
$\sin^2\theta\approx 0.23.$ 
\end{center} 

Let us perform the concrete estimates for the electron-nucleon interaction. 
In this case the induced electron-quark operator is
\begin{eqnarray}\label{eN}
\frac{G}{\sqrt{2}}\,\frac{\alpha}{3\pi}\,\log\frac{\Lambda^2}{M^2}\,v\,
\{\,q_{eq}\,[\,\frac{m}{m}\,\bar e \,i
\gamma_5\sigma_{\mu\nu}e\,\bar q\,\sigma_{\mu\nu}q\,-\,
\frac{1}{2m}\,\bar e\,i\gamma_5
\sigma_{\mu\nu}(p'_1-p_1)_\nu e\,\bar q\,\gamma_\mu q\,]\nonumber\\
                                                     +q_{qe}\,\frac{m_e}{m}
\,[\,(1-2a)\,\bar e\,i\gamma_5 e\,\bar q\, q\, -\, 3\,\bar e\,e\, 
\bar q\, i\gamma_5 q\,]\,\}.
\end{eqnarray}
Here $m$ and $m_{e}$ are the quark and electron masses, $v$ and $a$
are the quark vector and axial charges, $q_e$ and $q$ are the
dimensionless constants in the $T$-odd, $P$-even operators with the
explicit momenta belonging to electrons and quarks, respectively. We
have neglected here the contribution proportional to the electron vector
charge $v_{e}$ which is numerically small (see (\ref{ch})).  Operator
(\ref{eN}) should be summed over $u$- and $d$-quarks, and its
expectation value should be taken first over a nucleon and then over
a nucleus.

In the static approximation for nucleons, the only term in (\ref{eN})
that depends on both electron and nucleon spin is
$\gamma_5\sigma_{\mu\nu}\times\sigma_{\mu\nu}$. The dimensional
estimate for the nucleon expectation value of the operator $\bar
q\sigma_{\mu\nu} q$ is $\bar N\sigma_{\mu\nu}N$. Then the
dimensionless effective constant of the $T$- and $P$-odd tensor
electron-neutron interaction is
\begin{equation}
k_2 =\,\frac{\alpha}{3\pi}\,\log\frac{\Lambda^2}{M^2}\,
\frac{m_d}{2m}\,v_{d} q_{eq}\,\sim\, 10^{-6} q_{eq}.
\end{equation}
The upper limit on the constant $k_2$  obtained in \cite{lam} leads to the following
result:
\begin{equation}
q_{eq}\,<\,10^{-2}.
\end{equation}
For the $d$-quark mass we assume here the value $m_{d}\,=\,7$ MeV. 

Upper limits on the level 0.1 -- 1 can be extracted in this way from atomic
experiments for the constants $q_{qe}$ referring to the
electron-quark interaction with the derivatives in the quark vertex.

The analogous estimates for various quark-quark constants, as derived from
the results both for the atomic and neutron EDMs, lead to the upper
limits 
\begin{equation}\label{es1}
q_{qq}\,<\,1.
\end{equation}
We do not go here into details since in the next subsection much better
upper limits will be obtained for all these constants.

\subsection{TOPE Fermion-Fermion Interactions. \newline 
Two-Loop Approach}

The idea of the next improvement of the upper limits on the constants 
discussed \cite{tkh2,tkh3}, can be conveniently explained for the
case of hadrons. The previous advance from (\ref{cru}) to (\ref{es1})
was obtained by going over from the long-distance effects of the
usual weak interaction to the short-distance ones. It allowed us to
get rid of one small factor $Gm^2_\pi$ in formula (\ref{wh}), trading
it for $\alpha/\pi$ with some extra chiral suppression. But can we
get rid of the second factor $Gm^2_\pi$ in that formula? The answer
is: yes, we can.  Up to now we computed one-loop radiative correction
which transformed the interaction discussed into $T$- and $P$-odd
effective operator. Then we estimated in fact the long-distance
contribution of this operator to the neutron dipole moment. Now we
are going to make the next step: to calculate a completely
short-distance two-loop contribution of the TOPE interaction times
the weak interaction, directly to the quark EDM. And the latter at
least does not significantly exceed the neutron dipole moment. We gain in
this way even more than at the first step since now there is no more
chiral suppression factor $m_q/m_p$.

To regularize, at least partly, 
the related Feynman integrals, it is convenient
to introduce explicitly an axial boson of mass $\mu$ mediating the
TOPE interaction. It was pointed out long ago \cite{sud} that the
amplitudes (\ref{da}) could arise through the exchange by a neutral
pseudovector boson if its vertices contain the mixture of the
``normal" axial operator $\gamma_\mu\gamma_5$ and the ``anomalous" one
$i\gamma_5\sigma_{\mu\nu}(p'-p)_\nu$ of opposite CP-parity. We
consider however the introduction of this boson only as a convenient
way to soften the ultraviolet divergence of the integral from the
quadratic to logarithmic one, without any serious discussion of this
particle by itself. Thus the TOPE fermion-fermion amplitude can be
presented as
\begin{equation}
\frac{4\pi\beta}{\mu^2-k^2}\,\frac{1}{2m}\,\bar{\psi}_1i\gamma_5
\sigma_{\mu\nu}(p'_1-p_1)_\nu\psi_1\,\bar{\psi}_2\gamma_\mu\gamma_5 \psi_2.
\end{equation}
Here $\beta$ is the dimensionless coupling constant analogous to
$\alpha=1/137$ in QED.
Consider now the contribution of the two-loop diagram 1a to the fermion 
EDM. Here the dashed line represents the propagator of the axial 
boson $X$. The $Z$-boson exchange (the wavy line) introduces the parity 
nonconservation necessary to induce the EDM.

It can be easily seen that the contribution proportional to the mass of 
fermion 1, is small, $\sim G$. Therefore, this fermion  
will be taken as massless. Since the dipole moment interaction changes the 
fermion chirality, the lower vertex  of the $X$-line should be 
$i\gamma_5\sigma_{\mu\lambda}k_{\lambda}$ and the upper one, 
correspondingly, $\gamma_{\mu}\gamma_5$. Then according to the Furry 
theorem for the fermion loop, the upper vertex of the $Z$ line is 
$\gamma_{\nu}$, and the lower one, respectively, $\gamma_{\nu}\gamma_5$.
To simplify the calculations and result, we will neglect the mass of the 
fermion propagating in the upper loop as well (though for the heavy 
$t$-quark there are no special reasons to do so). 
Then the expression for the fermion loop 
is \cite{kh1}
\begin{equation}\label{anom}
\frac{i}{8\pi^2}F_{\alpha\beta}[\,\epsilon_{\alpha\beta\mu\nu}+
\frac{1}{k^2}\epsilon_{\alpha\beta\kappa\lambda}k_\kappa
(\delta_{\mu\lambda}k_{\nu}+\delta_{\nu\lambda}k_{\mu})\,];
\end{equation}
here $F_{\alpha\beta}$ is the strength of the external electromagnetic field.

When considering the lower (Compton) block of diagram 1a, one should also 
include the contact term
\begin{equation}\label{cont}
\frac{e}{2\sin \theta \cos \theta}\; 2\, a \,i\,\sigma_{\mu\nu},
\end{equation}
originating from the vertex 
\[i\gamma_5\sigma_{\mu\lambda}(p'-p)_{\lambda}\]
via the substitution 
\[p_{\mu}\psi\rightarrow[p_{\mu}-\frac{e}{2\sin \theta \cos \theta}\;
(v+a\gamma_5)\,Z_{\mu}]\,\psi\]
which makes this vertex gauge invariant with respect to the
$Z$-field. In particular, the inclusion of the contact vertex
(\ref{cont}) into the $XZ$ Compton scattering amplitude makes the
result of the calculation independent of the term
$k_{\nu}k_{\sigma}/M^2$ in the propagator of the $Z$-boson. Simple
calculations give now the following result for this contribution to the
EDM $d$ of fermion 1:
\begin{equation}\label{2l}
\frac{d}{e}\,=\,-\,\frac{\alpha\,\beta\, Q_2 a_1 v_2}{3\pi^2 m}\,
\log \frac{\Lambda^2}{M^2_{>}}.
\end{equation}
This formula refers to the general case in which fermion 2
propagating in the loop differs from the fermion 1 propagating in the
lower line. In particular, $a_1$ is the axial weak charge of the
first fermion, $v_2$ is the vector weak charge of the second fermion,
and $Q_2$ is its electric charge in the units of $e$. The logarithmic
dependence on the cut-off parameter $\Lambda$ is the result of the
nonrenormalizable coupling of the $X$-boson to the vertex with
derivatives $i\gamma_5\sigma_{\mu\rho}k_{\rho}$.  Although the result
(\ref{2l}) is presented with logarithmic accuracy, in all our
numerical estimates we will conservatively assume $\log
\Lambda^2/M^2_{>}$ to be of the order of unity ($M_{>}$ is the
largest of the masses $\mu$ and $M$). Let us emphasize again the
gauge invariance of this result with respect to the $Z$-field. As to
its gauge invariance with respect to the electromagnetic field, it is
self-evident from expression (\ref{anom}).

The contribution of diagram 1b
\begin{equation}
\frac{d_b}{e}\,=\,\frac{\alpha\,\beta\, Q_1 a_2 v_1}{36\pi^2 m}\,
\log^2\frac{\Lambda^2}{M^2_{>}}
\end{equation}
is much smaller numerically and can be neglected.

In the case of identical fermions there is also the contribution to the 
EDM of diagram 1c, but we will neglect it in our estimates with the
expectation that the result will not be grossly affected. One might
expect here that in the local limit, $\mu\rightarrow\infty$, the
effect for identical fermions 1 and 2 should vanish, which would
correspond to exact cancellation of diagrams 1a and 1c. However,
even for $\mu\gg M$ we can, to logarithmic accuracy, restrict the
integration over $k$ to $k\gg\mu$, where the TOPE interaction of
identical fermions is in no way a local one and therefore no
cancellation takes place.  

A consistent, gauge-independent calculation of the $W$-boson exchange
contribution to the induced EDM is again much more model-dependent
and we will not discuss it.  The same arguments as in the previous
section lead us to expect that the $Z$-boson contribution 
alone serves as
a conservative estimate for the induced EDM.

We will start the application of the general result (\ref{2l}) from
the case of the electron-electron TOPE interaction. Substituting into
formula (\ref{2l}) the numerical values (\ref{ch}) for $a_{e}$,
$v_{e}$, as well as $Q_{e}=-1$, we get
\begin{equation}
\frac{d_{e}}{e}\sim\beta_{ee}\cdot 10^{-19}\;\mbox{cm}.
\end{equation}
The experimental upper limit on the electron EDM \cite{com,hun} 
\begin{equation}\label{eedm}
\frac{d_{e}}{e}\,<\,10^{-26}\;\mbox{cm}.
\end{equation}
leads to the following
result for the constant $\beta_{ee}$ of the electron-electron TOPE
interaction:
\begin{equation}
\beta_{ee}\,<\,10^{-7}.
\end{equation}

In the same way we can get very strict upper limits on the
electron-nucleon and nucleon-nucleon TOPE interactions. The axial
charge of a fermion is always (up to a sign) $1/2$ and for any quark,
independently of its sort, the product $Qv$ is numerically close to
$1/9$. Then, using the experimental upper limit (\ref{nedm}) on the
neutron EDM and assuming for dimensional reasons that the neutron
dipole moment induced by the quark EDM is of about the same magnitude
as the latter, we get for $\beta_{qe}$, the TOPE quark-electron
interaction constant with derivatives in the quark vertex, the limit
\begin{equation}\label{qe}
\beta_{qe}\,<\,10^{-6}.
\end{equation}
For another electron-quark constant $\beta_{eq}$ (with the derivative
in the electron vertex) the constraint (\ref{eedm}) on the electron EDM
gives the upper limit
\begin{equation}\label{eq}
\beta_{eq}\,<\,3\cdot 10^{-8}. 
\end{equation}
For all quark-quark constants $\beta_{qq}$ the limit (\ref{nedm}) on the
neutron EDM gives
\begin{equation}\label{qq}
\beta_{qq}\,<\,3\cdot 10^{-7}.
\end{equation}
The latter refers as well to the ``coloured" TOPE interaction, with
the $SU(3)$ generators $t^a$ in each vertex. In this case the external
field on diagram 1 should not be electromagnetic, but rather a gluon
one. Again, for dimensional reasons, the neutron EDM induced in this
way should be of the same order of magnitude as the chromoelectric
dipole moment described by this diagram.

The constants $\beta$ introduced here, are related as follows to the 
constants $q$ used above:
\begin{equation}
\frac{4\pi\beta}{\mu^2}\,=\,\frac{G}{\sqrt 2}\,q,
\end{equation}
or
\begin{equation}
q\,=\,4\pi\beta\,\sqrt 2\,\left(\frac{m_p}{\mu}\right)^2 \cdot 10^5\,
=\,1.8\cdot 10^6\,\left(\frac{m_p}{\mu}\right)^2 \beta.
\end{equation}
Thus, the upper limits corresponding to (\ref{qe}), (\ref{eq}) and
(\ref{qq}) are
\begin{equation}
q_{qe}\,<\,2\left(\frac{m_p}{\mu}\right)^2, \;\;\;  
q_{eq}\,<\,0.05\left(\frac{m_p}{\mu}\right)^2, 
\;\;\;   q_{qq}\,<\,0.5\left(\frac{m_p}{\mu}\right)^2,
\end{equation}
respectively.

The upper limits on the constants $q$ in the interval 
$\,q\,<\,10^{-2}-\,1$, were
derived in the previous section under the assumption $\mu\geq M\sim
100\,m_p$. Under the same assumption, the limits we obtain here are
much better:
\begin{equation}\label{qlim}
q_{qe}\,<\,10^{-4}, \;\;\;\;q_{eq}\,<\,10^{-5}, \;\;\;\;q_{qq}\,<\,10^{-4}.
\end{equation}

Let us come back to the explanation of this gain. In the transition
from the effective four-fermion $T$- and $P$-odd operators obtained
in the previous section to the neutron EDM, we used the usual
hadronic scale of 1 GeV. But here the transition takes place on a
much higher scale of 100 GeV.

\subsection{Electron-Nucleon and Nucleon-Nucleon \newline 
Interactions. Conclusions}

Now, having obtained the above limits on the TOPE electron-quark and 
quark-quark interactions, what can we say about the corresponding 
electron-nucleon and nucleon-nucleon interactions? 

The answer for the electron-nucleon interaction is quite
straightforward one. Simple dimensional arguments lead to the
following estimates for the nucleon expectation values of the
relevant quark operators (the second one has been already mentioned):
\begin{equation}
\langle N|\,\bar{q}\gamma_{\mu}\gamma_5 q\,|N \rangle\,\sim\, 
\bar{N}\gamma_{\mu}\gamma_5 N,\;\;
\langle N|\,\bar{q}\gamma_5\sigma_{\mu\nu} q\,|N>\,\sim \,
\bar{N}\gamma_5\sigma_{\mu\nu} N.
\end{equation}
Therefore, the limits (\ref{qlim}) for $q_{qe,eq}$ are readily
translated into those for the constants of TOPE electron-nucleon
interactions:
\begin{equation}
q_{Ne}\,<\,10^{-4},  \;\;\;\;\;\;         q_{eN}\,<\,10^{-5}.
\end{equation}

Let us address now the nucleon-nucleon interactions. Note first of
all that in contrast to $T$- and $P$-odd nuclear forces, TOPE ones
cannot be mediated by $\pi^0$-meson exchange \cite{sim}. Indeed,
looking at the classification of the particle-antiparticle states in
the annihilation channel presented in Section 2.3, we see that at
$j=0$ the state 2 just does not exist.

The absence of this exchange can be attributed also to vanishing of a
TOPE $\pi^{0} NN$ vertex. As to the TOPE $\pi^{\pm} NN$ coupling, being 
hermitian, it should be written as
\begin{equation}
\bar p\,\gamma_5 n\,\pi^+ - \bar n\,\gamma_5 p\,\pi^-.
\end{equation}
This coupling does not lead to TOPE NN scattering amplitude in the
one-boson exchange approximation since after the interchange of this
vertex and of the strong one, the corresponding diagrams cancel.
TOPE one-boson exchange starts therefore with vector and pseudovector
bosons. Being mediated by heavier particles, the effective NN
interaction is further suppressed as compared to simple
estimates.

On the other hand, it follows already from general formulae
(\ref{da}) that a TOPE nucleon-nucleon scattering amplitude contains
an extra power of $p/{m}$ as compared to the usual $P$-odd weak
interaction. This means an extra suppression of roughly by an order
of magnitude as compared with the mentioned naive estimate
$G\,m^2_{\pi}\,q$.

Thus, even taking into account all the uncertainties of our
estimates, one can state that the relative strength of the TOPE
nuclear forces does not exceed $10^{-4}Gm^2_{\pi}$, or 
\begin{equation}
\alpha_T\,<\,10^{-11}.
\end{equation}

Various objections to this conclusion have died away, being withdrawn
implicitly or explicitly. The only one still worth mentioning is the
possibility that the contributions of various particles to the
fermion loop in diagram 1a cancel out. However, this possibility
emphasized in \cite{hhm} refers obviously to any estimates
(including those of \cite{hhm}) made in the absense of a
reliable theory. As to the analogy with the
well-known GIM mechanism mentioned in \cite{hhm}, it does not seem
relevant here.  The reasons for the GIM cancellation in the standard
model are well-known, but they also seem irrelevant to the issue of
nonrenormalizable TOPE interactions. On the other hand, too strong
cancellation here looks especially unlikely due to the large mass of
the $t$-quark. Moreover, a 
cancellation at the level of $10^{-6}$, which is required to change 
the upper limit $10^{-11}$,
set in \cite{tkh3}, to $10^{-5}$, as discussed in \cite{hhm},
seems quite improbable.

\section*{Acknowledgments}

The work was supported by the
Russian Foundation for Basic Research through grant No.95-02-04436-a.
The text was written during the visit to Vancouver. I truly appreciate
the kind hospitality extended to me at TRIUMF and UBC.


\bigskip

\begin{thebibliography}{99}

\bibitem{sf} O.P. Sushkov and V.V. Flambaum, Pis'ma 
Zh.Eksp.Teor.Fiz. 32, 377 (1980) [Sov.Phys.JETP Lett. 32,
353 (1980)].

\bibitem{fb} C.M. Frankle, J.D. Bowman, J.E. Bush, P.P.J. Delheij,
C.R. Gould, D.G. Haase {\it et al.}, 
Phys.Rev.Lett. 67, 564 (1991).

\bibitem{fb1} C.M. Frankle, J.D. Bowman, J.E. Bush, P.P.J. Delheij,
C.R. Gould, D.G. Haase {\it et al.},    
Phys.Rev. C 46, 778 (1992).

\bibitem{bg} J.D. Bowman, G.E. Garvey, C.R. Gould, A. Hayes and M.B.
Johnson, Phys.Rev.Lett. 68, 780 (1992).

\bibitem{kj} S.E. Koonin, C.W. Johnson and P. Vogel,
Phys.Rev.Lett. 69, 1163 (1992).

\bibitem{ab} N. Auerbach and J.D. Bowman, 
Phys.Rev. C 46, 2582 (1992).

\bibitem{lw} C.H. Lewenkopf and H.A. Weidenm\"{u}ller, 
Phys.Rev. C 46, 2601 (1992).

\bibitem{ts} L.V. Inzhechik, E.V. Mel'nikov, A.S. Khlebnikov, V.G.
Tsinoev and B.J. Ragozev, Zh.Eksp.Teor.Fiz. 93, 800 (1987) 
[Sov.Phys. JETP 66, 450 (1987)]; Yad.Fiz. 44,
1370 (1986) [Sov.J.Nucl.Phys. 44, 890 (1986)].

\bibitem{ts1} L.V. Inzhechik, A.S. Khlebnikov, V.G.
Tsinoev, B.J. Ragozev, M.Yu. Silin and Yu.M. Pen'kov, 
Zh.Eksp.Teor.Fiz. 93, 1560 (1987) [Sov.Phys. JETP 66, 897 (1987)].

\bibitem{sb} J.J. Szymanski, J.D. Bowman, M. Leuschner, B.A. Brown
and I.C. Girit, Phys.Rev. C 49, 3297 (1994).

\bibitem{sb1} J.J. Szymanski, J.D. Bowman, M. Leuschner, B.A. Brown
and I.C. Girit, Phys.Rev. C 52, 1713 (1995).

\bibitem{dkt1} V.F. Dmitriev, I.B. Khriplovich and V.B. Telitsin,
Phys.Rev. C 52, 1711 (1995).

\bibitem{mvm} D.M. Meekhof, P. Vetter, P.K. Majumder, S.K. Lamoreaux
and E.N. Fortson, Phys.Rev.Lett. 71, 3442 (1993).
 
\bibitem{fk} V.V. Flambaum and I.B. Khriplovich, 
Zh.Eksp.Teor.Fiz. 79, 1656 (1980) [Sov.Phys. JETP 52, 835 (1980)].

\bibitem{fks1} V.V. Flambaum, I.B. Khriplovich and O.P. Sushkov, 
Phys.Lett. B 145, 367 (1984).

\bibitem{khr} I.B. Khriplovich {\em Parity Nonconservation in Atomic
Phenomena} (Gordon and Breach, 1991).

\bibitem{dkt} V.F. Dmitriev, I.B. Khriplovich and V.B. Telitsin,
Nucl.Phys. A 577, 691 (1994).

\bibitem{dt} V.F. Dmitriev and V.B. Telitsin, to be published.

\bibitem{nsfk} V.N. Novikov, O.P. Sushkov, V.V. Flambaum and 
I.B. Khriplovich,
Zh.Eksp.Teor.Fiz. 73, 802 (1977) 
[Sov.Phys. JETP 46, 420 (1977)].

\bibitem{k} I.B. Khriplovich, Phys.Lett. A 197, 316 (1995).

\bibitem{wi} M.S. Noecker, B.P. Masterson, C.E. Wieman, 
Phys.Rev.Lett. 61, 310 (1988).

\bibitem{baird} N.H. Edwards, S.J. Phippp, P.E.G. Baird
and S.Nakayama, Phys.Rev.Lett. 74, 2654 (1995).

\bibitem{la} P.A. Vetter, D.M. Meekhof, P.K. Majumder, S.K. Lamoreaux
and E. N. Fortson, Phys.Rev.Lett. 74, 2658 (1995).

\bibitem{mzw} M.J.D. Macpherson, K.R. Zetie, R.B. Warrington, D.N. Stacey
and J.P. Hoare, Phys.Rev.Lett. 67, 2784 (1991).

\bibitem{lam} J.P. Jacobs, W.M. Klipstein, S.K. Lamoreaux, B.R. Heckel 
and E.N. Fortson, Phys.Rev. A 52, 3521 (1995).

\bibitem{sfk} O.P. Sushkov, V.V. Flambaum and I.B. Khriplovich,
Zh.Eksp.Teor.Fiz. 87, 1521 (1984) 
[Sov.Phys.JETP 60, 873 (1984)].

\bibitem{hh1} W.C. Haxton and E.M. Henley, Phys.Rev.Lett. 51, 1937 
(1983).

\bibitem{fks} V.V. Flambaum, I.B. Khriplovich and O.P. Sushkov, 
Nucl.Phys. A 449, 750 (1986).

\bibitem{amar} A.-M. M\aa rtensson-Pendrill, Phys.Rev.Lett. 54, 
1153 (1985).

\bibitem{kky} V.M. Khatsymovsky, I.B. Khriplovich and A.S. Yelkhovsky, 
Ann.Phys. 186, 1 (1988).

\bibitem{gw} G.F. Gunion and D. Wyler, Phys.Lett. B 248, 170
(1990).

\bibitem{cky} D. Chang, W.-Y. Keung and T.C. Yuan, Phys.Lett. B 251, 
608 (1990).

\bibitem{kh1u} I.B. Khriplovich, preprint BINP 96-16.

\bibitem{khz} I.B. Khriplovich and K.N. Zyablyuk, preprint BINP 96-13.

\bibitem{chk} N.K. Cheung, H.E. Henrikson and F. Boehm, 
Phys.Rev. C 16, 2381 (1977).

\bibitem{blh} J. Bystricky, F. Leah and P. Winternitz, 
J.Phys. 45, 207 (1984).

\bibitem{dav} C.A. Davies et al., Phys.Rev. C 33, 1196 (1986).

\bibitem{fps} J.B. French, A. Pandey and J. Smith, in {\em Tests of Time 
Reversal Invariance in Neutron Physics}, ed. N.R.  Roberson, C.R.
Gould and J.D. Bowman (World Scientific, Singapore, 1987).

\bibitem{porkoz} M.G. Kozlov and S.G. Porsev, Phys.Lett. A 142, 
233 (1989); Yad.Fiz. 51, 1056 (1990) 
[Sov.J.Nucl.Phys. 51 (1990)].

\bibitem{mospor} A.N. Moskalev and S.G. Porsev, Yad.Fiz. 49,
1266 (1989) [Sov.J.Nucl.Phys. 49, (1989)].

\bibitem{cont} R.S. Conti, in {\em Time Reversal -- the Arthur Rich
Memorial Symposium}, eds. M. Skalsey, P. Bucksbaum, R.S. Conti and
D.W. Gidley (AIP, New York, 1991).

\bibitem{chr} R.S. Conti, S. Hatamian and A. Rich, Phys.Rev. A 33, 
3495 (1986).

\bibitem{hksw} P. Herczeg, J. Kambor, M. Simonius and D. Wyler, to be
published.

\bibitem{wol} L. Wolfenstein, Nucl.Phys. B 77, 375 (1974).

\bibitem{her} P. Herczeg, in {\em Tests of Time Reversal Invariance
in Neutron Physics}, ed. N.R.Roberson, C.R. Gould and J.D. Bowman
(World Scientific, Singapore, 1987).

\bibitem{tkh1} I.B. Khriplovich, Nucl.Phys. B 352, 385 (1991).

\bibitem{sm} K.F. Smith {\it et al}, Phys.Lett. B, 234, 191 (1990).

\bibitem{al} I.S. Altarev {\it et al}, Phys.Lett. B 276,
242 (1992).

\bibitem{kozl} M.G. Kozlov, unpublished.

\bibitem{khrip} I.B. Khriplovich, unpublished.

\bibitem{step} E. Stephens, Ph.D. thesis, University of Oxford, 1992,
unpublished.

\bibitem{hh} W.C. Haxton and A. H\"{o}ring, Nucl.Phys. A 560, 469 
(1993). 

\bibitem{hhm} W.C. Haxton, A. H\"{o}ring and M. Musolf, 
Phys.Rev. D 50, 3422 (1994). 

\bibitem{egh} J. Engel, C.R. Gould and V. Hnizdo, Phys.Rev.Lett. 73, 
3508 (1994).
                                                                                                                
\bibitem{vor} O.K. Vorov, to be published.

\bibitem{efs} J. Engel, P.H. Frampton and R.P. Springer, 
to be published.

\bibitem{tkh4} I.B. Khriplovich, in {\em Time Reversal -- the Arthur
Rich Memorial Symposium}, eds. M. Skalsey, P. Bucksbaum, R.S. Conti
and D.W. Gidley (AIP, New York, 1991). 

\bibitem{tkh2} I.B. Khriplovich, Pis'ma Zh.Eksp.Teor.Fiz. 52, 
1065 (1990) [Sov.Phys.JETP Lett. 52, 461 (1990)].

\bibitem{tkh3} R.S. Conti and I.B. Khriplovich, Phys.Rev.Lett. 68, 
3262 (1992).
 
\bibitem{sud} E.C.G. Sudarshan, Proc.Roy.Soc. London A 305, 319 (1968).

\bibitem{kh1} I.B. Khriplovich, Yad.Fiz. 44, 1019 (1986)
[Sov.J.Nucl.Phys. 44, 659 (1986)].

\bibitem{com} K. Abdullah, C. Carlberg, E.D. Commins, H. Gould and
S.B. Ross, Phys.Rev.Lett. 65, 2347 (1990).

\bibitem{hun} S.A. Murthy, D. Krause, Jr., Z.L. Li and L.R. Hunter, 
Phys.Rev.Lett. 63, 965 (1989).

\bibitem{sim} M. Simonius, Phys.Lett. B 58, 147 (1975).




\end{thebibliography}
\end{document}